\documentclass[pra,10pt, twocolumn, aps, superscriptaddress, showpacs, groupaddress]{revtex4-1}
\usepackage[T1]{fontenc}
\usepackage{bigints}
\usepackage{amssymb}
\usepackage{graphicx}
\usepackage{amsmath}
\usepackage{dcolumn}
\usepackage{multirow}
\usepackage{hyperref}
\usepackage[normalem]{ulem}
\usepackage{physics}
\usepackage{bm}
\usepackage{comment}
\usepackage{xcolor}
\usepackage[caption=false]{subfig}
\usepackage{siunitx}
\usepackage{dsfont}
\usepackage{bbold}
\usepackage[toc]{appendix}
\usepackage{relsize}
\usepackage{leftidx}
\usepackage{bbm}
\usepackage{stackengine}
\usepackage{float}
\usepackage{tikz}
\usetikzlibrary{arrows.meta,calc,decorations.pathmorphing,positioning,fit,backgrounds}
\hypersetup{colorlinks=true,linkcolor=blue,linktoc=page, citecolor=blue,urlcolor=blue}

\everymath{\displaystyle} 

\begin{document}

\title{Negative diffusivity of excitons in electron-hole plasmas}

\author{H. Ter\c{c}as}
\affiliation{Instituto Superior de Engenharia de Lisboa,
  Instituto Polit\'{e}cnico de Lisboa,
  Rua Conselheiro Em\'{i}dio Navarro, 1959-007 Lisboa, Portugal}
\affiliation{GoLP/Instituto de Plasmas e Fus\~{a}o Nuclear,
  Instituto Superior T\'{e}cnico, Universidade de Lisboa,
  1049-001 Lisboa, Portugal}

\author{V. N. Mantsevich}
\affiliation{Physics Department,
Lomonosov Moscow State University, 119991 Moscow, Russia}

\begin{abstract}
We develop a minimal hydrodynamic framework to describe exciton transport in the presence of an electron–hole plasma in two‑dimensional semiconductors. Treating excitons, electrons, and holes as coupled fluids, we show that exciton diffusion is strongly renormalized by momentum exchange with the plasma. In the collisional regime, mutual diffusion leads to a nontrivial redistribution of transport coefficients but preserves the positivity of the exciton diffusivity. In contrast, when plasma inertia and collective charge oscillations are accounted for, the exciton diffusive mode hybridizes with acoustic plasma modes, giving rise to a dynamical instability manifested as an effective negative diffusion coefficient. We demonstrate that this instability originates from the non‑equilibrium coupling between slow excitons and fast plasma degrees of freedom, rather than from nonlinear diffusion or thermodynamic effects. Our results provide a unified physical mechanism for negative exciton diffusivity reported in recent experiments and establish collective plasma dynamics as a key control parameter of exciton transport in two‑dimensional materials.
\end{abstract}

\maketitle

\section{Introduction}

Excitons are bound electron–hole pairs arising from the Coulomb interaction in semiconductors \cite{Frenkel1931}. Under optical excitation, excitons are formed in a wide variety of material platforms, including inorganic and organic molecular semiconductors \cite{Colby2010,Berghuis2021}, colloidal quantum dots and nanoplatelets \cite{Klimov2000,Ithurria2011,Smirnov2019,Rabouw2016,Olutas2015,Brumberg2019}, semiconductor nanostructures such as quantum wells and quantum wires \cite{Scholes2006,Grim2014}, metal-halide perovskites \cite{Belykh2019,Ziegler2020,Magdaleno2021,Seitz2020}, and atomically thin transition metal dichalcogenides (TMDs) \cite{Chernikov2014,Wang2018}.

Two-dimensional materials are of particular interest in this context, as reduced dielectric screening leads to large exciton binding energies, enabling robust excitonic phenomena even at room temperature. Beyond their fundamental relevance to optical properties, excitons play an active role in energy and information transport, as they can propagate over mesoscopic length scales within the material plane \cite{Hillmer1988,Steininger1996,Rapaport2004,Kumar2014,Kato2016}. This property makes excitons appealing candidates for signal carriers in optoelectronic and chip-scale photonic architectures \cite{Baldo2009,Causin2022}. The combination of rich many-body physics and technological potential has therefore positioned exciton transport in two-dimensional systems as one of the most active and rapidly developing research directions in contemporary condensed-matter physics.

Experimental studies of exciton transport in two-dimensional materials have revealed predominantly diffusive propagation under a wide range of conditions \cite{Kumar2014,Yuan2017}. Key parameters such as the exciton diffusion coefficient and diffusion length are strongly influenced by disorder \cite{Vlaming2013,Kurilovich2020,Kurilovich2022,Lee2015} and by inter-particle interactions. Among the latter, exciton–exciton interactions \cite{Uddin2020,Cheng2021,Wagner2023} and exciton–phonon coupling \cite{Glazov2019} play a central role. Experimentally, exciton diffusion is commonly inferred from optical techniques, including pump-intensity-dependent photoluminescence \cite{Shaw2008,Stevens2001} and spatially and time-resolved microphotoluminescence measurements \cite{Akselrod2014,Ginsberg2020}. From a theoretical perspective, exciton transport has been extensively studied using semiclassical approaches based on the Boltzmann kinetic equation \cite{Zipfel2020,Choi2023}, which can in some cases be extended to incorporate quantum interference effects \cite{Glazov2022}. Complementary microscopic descriptions rely on tight-binding models \cite{Kenkre1983,Heijs2005} or kinetic Monte Carlo simulations \cite{Akselrod2014_1,Miyazaki2012}.
A particularly intriguing phenomenon reported in recent years is the emergence of negative effective exciton diffusivity in two-dimensional systems \cite{Wietek2024,Kurilovich2023,Adejumobi2024}. Such behavior is generically absent in single-component diffusive systems and instead appears to be intrinsic to multicomponent scenarios. Experiments have demonstrated that excitons may split into sub-populations characterized by distinct transport properties, with exchange processes between these sub-populations giving rise to nonlinear or even negative effective diffusion \cite{Berghuis2021,Wietek2024,Kurilovich2023,Rosati2020,Rosati2021}. Physically, fast and slow exciton species can originate from bright and dark excitons \cite{Berghuis2021,Rosati2020,Rosati2021}, free and trapped excitons \cite{Lin2016,Kurilovich2022}, or from the coexistence of excitons with an electron–hole plasma \cite{Wietek2024}. Recent theoretical treatments have attributed negative 
\textcolor{blue}{diffusivity} either to nonlinear interaction effects \cite{Wietek2024} or to exchange processes between free and trapped exciton states \cite{Kurilovich2023}.
Beyond the diffusive regime, excitons may also exhibit collective transport behavior. Notable examples include exciton superfluidity \cite{Fogler2014} and hydrodynamic flow \cite{Mantsevich2024}. Experimental evidence for superfluid-like exciton transport has recently been reported in a monolayer $\mathrm{MoS_2}$ system \cite{Aguila2023}, underscoring the richness of non-diffusive exciton dynamics in two-dimensional materials.
In this work, we introduce a minimal theoretical framework to investigate exciton diffusion in the presence of an electron–hole plasma. Treating excitons, electrons, and holes as coupled fluids, we analyze how momentum exchange and collective plasma dynamics modify exciton transport. We first consider the diffusive regime in a quasi-neutral plasma, neglecting plasmons, and derive the corresponding effective diffusion equations. We then incorporate plasma inertia and collective charge oscillations, demonstrating the emergence of an exciton bubble instability associated with an effective negative diffusion coefficient. 

\section{Basic model}

We start by constructing the most basic, yet phenomenologically correct theory describing the diffusive processes of excitons in a two-dimensional electron-hole plasma. The goal is to provide a theoretical justification for the renormalization of the intrinsic exciton diffusion coefficient, $D_X = k_B T / (m_X \gamma_X)$, arising from the coupled dynamics of excitons, electrons, and holes, where $\gamma_X$ accounts for intrinsic processes such as finite lifetime and impurity scattering. We consider a two-dimensional semiconductor (e.g., a TMD) in which these three fluids coexist and interact. The exciton fluid is governed by the following equations
\begin{equation}
\begin{array}{l}
\frac{\partial n_X}{\partial t}+\bm \nabla\cdot\left(n_X {\bf u}_X\right)=0 \\\\
\left(\frac{D}{Dt} + \gamma_X \right){\bf u}_X=-k_B T_X\frac{\bm \nabla  n_X}{m_X n_X} -\sum_{\alpha=\{e,h\}}\gamma_\alpha \left( {\bf u}_X-{\bf u}_\alpha\right),
\end{array} 
\label{eq_exciton}
\end{equation}
where $D/Dt=\partial/\partial t+\bf v\cdot\bm \nabla$ is the Lagrange derivative, $\gamma_\alpha$ is the rate at which elastic collisions with the $\alpha-$th species take place. Conversely, the electrons/holes are described by the following equations
\begin{equation}
\begin{array}{l}
\frac{\partial n_\alpha}{\partial t}+\bm \nabla\cdot\left(n_\alpha {\bf u}_\alpha\right)=0 \\\\
\frac{D}{Dt}{\bf u}_\alpha=\frac{q_\alpha{\bf E}}{m_\alpha}-k_B T_\alpha\frac{\bm \nabla  n_\alpha}{m_\alpha n_\alpha} +\gamma_\alpha \left( {\bf u}_X-{\bf u}_\alpha\right),\\\\
\bm\nabla \cdot {\bf E}=\frac{e}{\epsilon_0}\left(\rho_h-\rho_e\right)=\frac{e}{\epsilon_0}\left(n_h-n_e\right)\delta(z),
\end{array}
\label{eq_plasma} 
\end{equation}
where the Dirac-delta $\delta(z)$ assures that the physics is happening in two dimensions. 

\section{Exciton diffusion in collisional plasmas}

As a first approach, we shall consider that quasi-neutrality is satisfied, $n_e\simeq n_h$, such that the Poisson Eq. in \eqref{eq_plasma} can be neglected. In an intermediate situation, the so-called {\it ambipolar regime}, an ambipolar electric field ${\bf E}_{\rm amb}$ is produced, keeping the electron-hole plasma moving together. However, this effect is particular prominent in heavy-hole systems, $m_h/m_e\gg 1$. While we may consider this variation quite straigthforwardly, for the purpose of the current estimations we assume the symmetric effective mass case, $m_e\sim m_h$, such that ${\bf E}_{\rm amb}\sim 0$. The situation of close effective masses of electrons and holes is realized, in particular, in Mo$-$based TMDs \cite{Kormnyos2015}. Since electrons and holes move as whole, it is therefore convenient to work with reduced variables, 
\begin{align}
n_p &=\frac{m_en_e+m_h n_h}{m_e+m_h}\simeq n_e +n_h, \nonumber \\ 
{\bf u}_p &=\frac{m_e{\bf u}_e+m_h {\bf u}_h}{m_e+m_h}\simeq \frac{{\bf u}_e +{\bf u}_h}{2}.
\end{align}
From Eq. \eqref{eq_plasma}, we may write
\begin{equation}
\begin{array}{l}
\frac{\partial n_p}{\partial t}+\bm \nabla\cdot\left(n_p {\bf u}_p\right)\simeq 0 \\\\
\frac{D}{Dt}{\bf u}_p\simeq -k_B T_p\frac{\bm \nabla  n_p}{m_p n_p} +\gamma_p \left( {\bf u}_X-{\bf u}_p\right).
\end{array}
\label{eq_plasma2}
\end{equation}
where $T_p=(m_en_e T_e+m_h n_h T_h)/(m_e n_e +m_h n_h)\simeq (T_e+T_h)/2$, is the plasma temperature, and $\gamma_p=(\gamma_e+\gamma_h)/2$. In that case, the exciton momentum conservation equation reads
\begin{equation}
\left(\frac{D}{Dt} + \gamma_X \right){\bf u}_x=-k_B T_X\frac{\bm \nabla  n_X}{m_X n_X} -\gamma_p \left( {\bf u}_X-{\bf u}_p\right).
\label{eq_exciton2}
\end{equation}
We now assume that the balanced gas condition, in which the exciton and plasma fractions are comparable, $\alpha_X\sim \alpha_p$. In that case, if one system behaves diffusively, the other will behave similarly. Mathematically, diffusive behavior is expected when the fluid varies slowly in the Lagrange frame, i.e. $D/Dt\ll \gamma_{X,p}$. As such, Eq. \eqref{eq_plasma2} provides 
\begin{equation}
{\bf u}_X^{(\rm diff)}\simeq {\bf u}_p^{(\rm diff)}+\frac{k_BT_p}{n_pm_p\gamma_p}\bm\nabla n_p. 
\end{equation}
Similarly, from Eq. \eqref{eq_exciton2}, the same procedure yields
\begin{equation}
{\bf u}_X^{(\rm diff)}\simeq \frac{\gamma_p}{\gamma_x+\gamma_p} {\bf u}_p^{(\rm diff)}-\frac{k_BT_X}{m_X n_X \left(\gamma_X+\gamma_p\right)}\bm\nabla n_X.
\label{eq_exciton3}
\end{equation}
Solving for both velocities, we get
\begin{align}
{\bf u}_p^{(\rm diff)} &\simeq -\frac{k_B T_X}{m_X n_X \gamma_X}\bm \nabla n_X -\frac{k_B T_p}{m_p n_p \widetilde \gamma}\bm \nabla n_p,  \nonumber \\[10 pt] {\bf u}_X^{(\rm diff)} &\simeq - \frac{k_B T_X}{m_X n_X \gamma_X}\bm \nabla n_X - \frac{k_B T_p}{m_p n_p\gamma_X}\bm \nabla n_p,
\end{align}
where $\widetilde \gamma=\gamma_X\gamma_p/(\gamma_X +\gamma_p)$ is the reduced collision rate. Putting in more explicit grounds, we get the diffusive two-fluid model
\begin{equation}
\begin{array}{l}
\frac{\partial n_X}{\partial t}- D_X\nabla^2 n_X- \bm\nabla \cdot \left(D_{Xp}\bm \nabla n_p \right)=0 , \\\\
\frac{\partial n_p}{ \partial t}- D_p\nabla^2 n_p - \bm\nabla \cdot \left(D_{pX}\bm \nabla n_X \right)=0, 
\end{array}
\label{eq_diffusion_total}
\end{equation}
where $D_X =k_B T_X/(m_X \gamma_X)$ and $D_p = k_B T_p/(m_p \widetilde \gamma)$ are the self-diffusion coefficients, and the $D_{Xp}$ and $D_{pX}$ are the mutual diffusion coefficients,

\begin{equation}
D_{Xp}=\frac{k_B T_X}{m_X\gamma_X}\frac{n_X}{n_p}, \quad  D_{pX}=\frac{k_B T_p}{m_p\gamma_X}\frac{n_p}{n_X}.
\end{equation}
\subsection{Dilute plasma case: adiabatic ellimination}

As a consistency check, we should try to observe if the plasma may, indeed, lead to negative diffusivity of excitons. To see that, let us adiabatically eliminate the plasma from the dynamics, assuming $\partial_t n_p \ll \partial_t n_X$. By doing that, we get $\bm \nabla n_p\simeq -(D_{pX}/D_p)\bm\nabla n_X$. Plugging back in the exciton diffusion coefficient in Eq. \eqref{eq_diffusion_total}, we obtain

\begin{equation}
\frac{\partial n_X}{\partial t}- \bm\nabla \cdot\left(\widetilde D_X\bm \nabla n_X \right)=0,
\end{equation}
where the effective excition diffusion coefficient is given by
\begin{equation}
\widetilde D_X= D_X-\frac{D_{pX}D_{Xp}}{D_p}=D_X\left(1-\frac{\gamma_p}{\gamma_p+\gamma_X}\right)>0. 
\label{eq:DeX_sym}
\end{equation}
This shows that any negative exciton diffusion must be balanced by positive plasma diffusion. Thus, negative diffusivity is restricted to the exciton subsystem, and the overall system remains positively diffusive.

\subsection{Dense plasma case: full self-consistent dynamics}

To retrieve such phenomenology, we should couple the plasma dynamics. Coming back to Eq. \eqref{eq_diffusion_total}, and performing perturbations around some equilibria, $n_j \simeq n_{j,0}+n_{j, 1}$, we get
\begin{equation}
\begin{array}{l}
\frac{\partial n_{X,1}}{\partial t}- D_X\nabla^2 n_{X,1}-\bm\nabla \cdot \left(D_{Xp}\bm \nabla n_{p,1} \right)=0 , \\\\
\frac{\partial n_{p,1}}{ \partial t}- D_p\nabla^2 n_{p,1}-\bm\nabla \cdot \left(D_{pX}\bm \nabla n_{X,1} \right)=0. 
\end{array}
\label{eq_diffusion_fourier}
\end{equation}
where the diffusion coefficients are intended to be evaluated at zero-th order, $D_{ij}[n_i, n_j]=D_{ij}[n_{i,0}, n_{j,0}]$. By performing a Fourier analysis, we get a secular equation of the form ${\rm det} M=0$, where $M$ reads
\begin{equation}
M=\left[\begin{array}{cc}
\omega+iD_X k^2 & -iD_{Xp} k^2 \\\\
-iD_{pX} k^2 & \omega+iD_p k^2
\end{array}
\right].
\end{equation}
Observing that $D_{Xp}D_{pX}=D_XD_P\widetilde \gamma/\gamma_X$, the density dependence is explicitly eliminated in the secular equation (although it may still exist implicitly via the dependence of the collision rates $\gamma_X$ and $\gamma_p$). Finally, the characteristic polynomial provides  
\begin{equation}
(\omega +iD_Xk^2)(\omega +i D_p k^2)- \frac{\gamma_p}{\gamma_p+\gamma_X}D_p D_Xk^4=0.
\end{equation}
In the absence of the collisions with the plasma, $\gamma_p\to 0$, the excitons exhibit a diffusive mode, $\omega=-iD_X k^2$, corresponding to pure damping. When the two systems are coupled, however, the dispersion relation reads
\begin{align}
\omega &=-\frac{ik^2}{2}\left[(D_X+D_p)\right.\nonumber \\ & \left.  \pm \sqrt{4\frac{\gamma_p}{\gamma_p+\gamma_X}D_p D_X +(D_X-D_p)^2}\right].
\label{eq_disp1}
\end{align}
Since $\gamma_p/(\gamma_p+\gamma_X)<1$, the imaginary part is always negative, meaning that negative diffusion is absent in diffusive plasmas. In what follows, we unlock an extra degree of freedom of the system, the plasmons, and show how they are at the origin of negative diffusivity of excitons. 

\subsection{Effect of mass asymmetry: the ambipolar correction}
\label{sec:ambipolar}

The symmetric-mass approximation $m_e \simeq m_h$ adopted above is
particularly well justified for Mo-based TMDs~\cite{Kormnyos2015}, but is less accurate in
W-based compounds (e.g.\ WSe$_2$ or WS$_2$) where the ratio
$m_h/m_e$ can differ from unity by up to $\sim 30\%$~\cite{Kormnyos2015}.
When $m_e \neq m_h$, the two carriers no longer drift at the same
speed in the absence of an external field, and charge separation
builds a Dember (ambipolar) electric field $\mathbf{E}_\text{amb}$
that enforces quasi-neutrality~\cite{vanRoosbroeck1953,Meyer1980}. To incorporate this effect within the present framework, we retain the
full Poisson equation for $\mathbf{E}$ and define the ambipolar velocity $\mathbf{u}_p^\text{amb}
 = -D_a \bm \nabla n_p/n_p $ and diffusion coefficient in the standard
way~\cite{Meyer1980,Nelson1982},
\begin{equation}
D_a = \frac{(D_e/m_e + D_h/m_h)\,m_e m_h}{m_e + m_h}\simeq \frac{2D_eD_h}{D_e+D_h},
\label{eq:Damb}
\end{equation}
where $D_e = k_BT_e/(m_e\gamma_e)$ and
$D_h = k_BT_h/(m_h\gamma_h)$ are the single-species diffusion coefficients, and the second equality in Eq.~\eqref{eq:Damb} holds for $T_e = T_h$ and $m_e \simeq m_h$. In the presence of mass asymmetry, the plasma continuity equation is modified: the self-diffusion coefficient $D_p$
is replaced by $D_a$, while the cross-diffusion term $D_{pX}$ acquires an additional prefactor
\begin{equation}
 \eta = \frac{2m_em_h}{(m_e+m_h)^2},
\label{eq:eta}
\end{equation}
which satisfies $\eta \leq 1$ (with $\eta = 1/2$ in the equal-mass limit). The effective exciton diffusion coefficient
then reads
\begin{equation}
D_X^\text{eff} = D_X\!\left(1 - \frac{\gamma_p}{\gamma_p+\gamma_X}\,\eta\right) > 0,
\label{eq:DXeff_amb}
\end{equation}
showing that mass asymmetry \emph{reduces} the magnitude of the plasma renormalization relative to the equal-mass case but does not qualitatively alter the physics.

\section{Exciton diffusion in collisionless plasmas: effect of the plasmons}

In what follows, we may be interested in the regime in which advection dominates over plasma diffusion, but still subdominant in respect to exciton diffusion, $\gamma_X\simeq D/Dt \gg\gamma_p$. In such a regime, let us analyze the coupled dynamics of the plasma and exciton subsystems. For such a description, we relax the quasi-neutrality condition in Eqs. \eqref{eq_plasma}, and conveniently use Poisson's equation in Fourier space, 

\begin{equation}
\widetilde\phi({\bf k})=i\frac{e}{2\varepsilon_{0}k}( \widetilde n_{h}-\widetilde n_{e}),
\label{eq_system3}
\end{equation}
with $n_{0e}=n_{0h}\equiv n_0$ and $\widetilde f(k)=\int d^2x f({\bf x})e^{i\bf k \cdot x}$ denoting the Fourier transform, which relates to the electric field as ${\bf} \widetilde E({\bf k})=i{\bf k} \widetilde\phi(k)$. 

\begin{figure}[t!]
\includegraphics[width=\columnwidth]{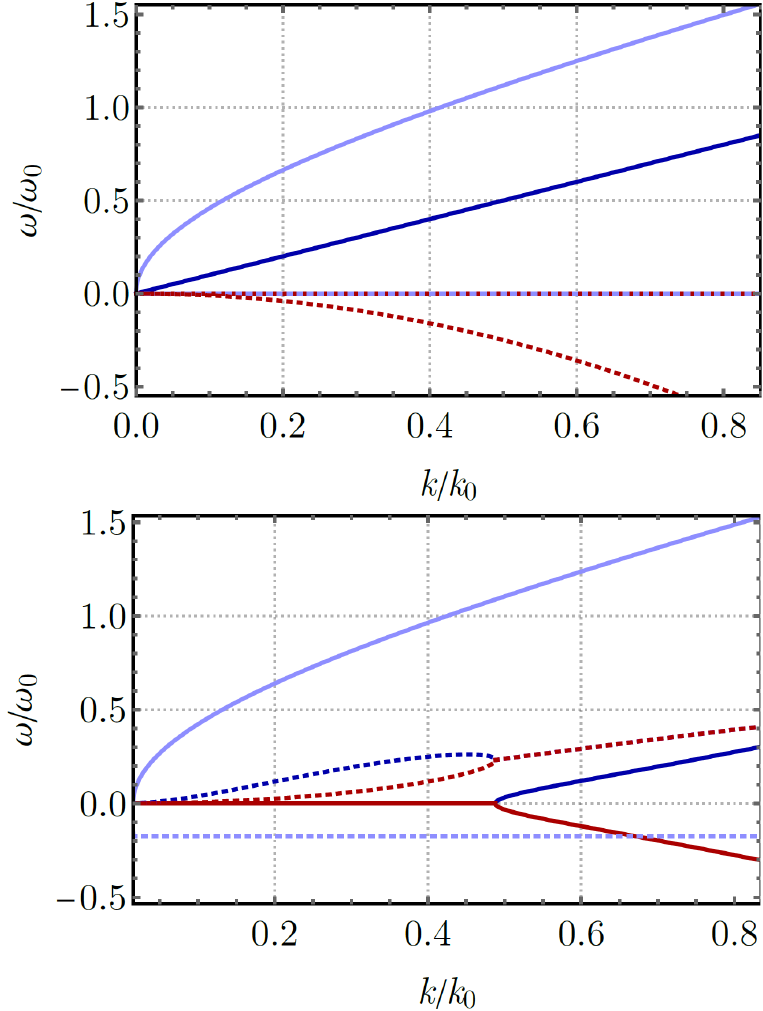}
\caption{Dispersion relation of the exciton-plasma system, in units of $k_0=e^2n_0/(2m_eD_X)$ and $\omega_0=D_Xk_0^2$. Top panel: normal modes in the absence of exciton-plasma coupling, $\gamma_e=\gamma_h=0$, featuring a diffusive mode ($\omega_1$, red line), plasma acoustic ($\omega_2$, light blue line) and plasma optical ($\omega_3$, dark blue line). Right panel: the same modes as obtained for $\gamma_e=\gamma_h=0.1\gamma_X$, and $\bar v_{X}=2.8 v_{\rm th}$. In all the cases, we have assumed a diffusion constant of $D_X=0.9v_{\rm th}^2/\omega_p$.}
\label{fig_dispersion}
\end{figure}

For excitons, we assume the diffusive limit $\partial_{t}\textbf{u}_X\ll\gamma_X\textbf{u}_X$, such that 
\begin{eqnarray}
\textbf{u}_X\simeq -\frac{v_X^{2}}{\gamma_X n_{0X}} \bm{\nabla}n_{X}+\sum_{\alpha}\frac{\gamma_{\alpha}}{\gamma_X}\textbf{u}_{\alpha}.
\label{eqn:one}
\end{eqnarray}
Substituting Eq. \eqref{eqn:one} to the continuity equation for excitons, one can get
\begin{eqnarray}
\frac{\partial n_X}{\partial t}-\bm{\nabla}\cdot \left(D_X\bm{\nabla}n_X\right)+\sum_{\alpha}\frac{\gamma_{\alpha}}{\gamma_X}\textbf{u}_{\alpha}=0.
\label{eqn:two}
\end{eqnarray}
Taking the second derivative of the electron/hole continuity equation, and putting Eqs. \eqref{eq_system3} and \eqref{eqn:one} together, one can eliminate $\textbf{u}_X$ from the dynamics of plasma subsystem, thus obtaining
\begin{align}
-\omega^2 n_{\alpha} &\pm\omega_{p}^{2}(n_{e}-n_{h})-i\gamma_{\alpha}\omega  n_{\alpha} +v_{\alpha}^{2}k^2n_{\alpha}\nonumber \\ &+i\gamma_{\alpha}n_{0\alpha}{\bf k}\cdot \left[-i\frac{v_{X}^{2}{\bf k}}{n_{0x}\gamma_X}n_{X}+\sum_{\alpha}\frac{\gamma_{\alpha}}{\gamma_X}\textbf{u}_{\alpha} \right]=0,
\label{eqn:three}
\end{align}
where $\omega_p=\sqrt{e^2n_0 k/(2\varepsilon_0 m_e )}$ is the 2D (dispersive) plasma frequency. Since $\gamma_{\alpha}/\gamma_X\ll 1$, the last term in Eq. \eqref{eqn:three} can be neglected, and together with Eq. \eqref{eqn:two}, yields the secular system
\begin{align}
\left[\omega^{2}-(\omega_{p}^{2}+v_{e}^{2}k^{2}+i\gamma_{e}\omega) \right]\widetilde{n}_{e}+\omega_{p}^{2}\widetilde{n}_{h}+\frac{\gamma_{e}}{\gamma_X}\bar v_X^{2}k^{2}\widetilde{n}_X=0,\nonumber\\
\left[\omega^{2}-\left(\omega_{p}^{2}+v_{h}^{2}k^{2}+i\gamma_{h}\omega \right)\right] \widetilde{n}_{h}+\omega_{p}^{2}\widetilde{n}_{e}+\frac{\gamma_{h}}{\gamma_X}\bar v_X^{2}k^{2}\widetilde{n}_X=0,\nonumber\\
-i\omega\widetilde{n}_X+D_X k^{2}\widetilde{n}_X+i \omega \frac{n_{0X}}{n_{0}}\left(\frac{\gamma_{e}}{\gamma_X}\widetilde{n}_{e}+\frac{\gamma_{h}}{\gamma_X}\widetilde{n}_{h}\right)=0,
\label{eq:secularTotal}
\end{align}
%
 with $\bar v_X=v_X\sqrt{n_{0}/n_{0X}}$ denoting the renormalized exction thermal speed. The dispersion relation contains three roots, which, in the absence of coupling ($\gamma_e=\gamma_h=0$), correspond to the bare modes $\omega={\omega_1,\omega_2,\omega_3}$, where
\begin{equation}
\omega_1=-iD_X k^2,  \quad \omega_2=v_{\rm th}k, \quad \omega_3=\sqrt{2\omega_p^2+v_{\rm th}^2k^2},
\end{equation}
with $v_{\rm th}\simeq (v_e+v_h)/2$, respectively corresponding to the exciton diffusive mode, and to the plasma acoustic and optical modes. For finite values of $\gamma_\alpha$, the bare modes hybridize. Interestingly, the diffusive mode mixes with the acoustic mode of the plasma and acquires a positive imaginary part, which is associated to a dynamical instability (or gain), as depicted in Fig. \ref{fig_dispersion}. Such a gain can be reinterpreted as the emergence of a negative diffusion coefficient.

To explicitly observe the effects of the plasma in the diffusion processes, we introduce the auxiliary functions
$A_e=\omega^{2}-(\omega_{p}^{2}+v_{e}^{2}k^{2}+i\gamma_{e}\omega)$ and $A_h=\omega^{2}-(\omega_{p}^{2}+v_{h}^{2}k^{2}+i\gamma_{h}\omega)$, such that the determinant of the secular matrix in Eq. \eqref{eq:secularTotal}
reads
\begin{align}
&\left(-i\omega + D_X k^2\right)\left(A_eA_h  -\omega_p^4\right)
\nonumber \\ &-i\omega\,\frac{\bar v_X^2k^2}{\gamma_X^2}
\left(
\gamma_e^2A_h+\gamma_h^2A_e-2\gamma_e\gamma_h\omega_p^2
\right)=0.
\label{eq:secular3x3}
\end{align}
This allows us to define an effective exciton diffusion coefficient in Fourier
space as
\begin{align}
\left[-i\omega+\widetilde D_X(k,\omega)k^2\right]\widetilde n_X=0,
\end{align}
with
\begin{equation}
\widetilde D_X(k,\omega)
=
D_X
-
i\omega\,\frac{\bar v_X^2}{\gamma_X^2}
\frac{
\gamma_e^2A_h+\gamma_h^2A_e-2\gamma_e\gamma_h\omega_p^2
}{
A_eA_h-\omega_p^4
}.
\label{eq:Deff_full}
\end{equation}
Equation~\eqref{eq:Deff_full} makes explicit that the exciton diffusivity is
renormalized by the full collective response of the electron--hole plasma,
including both acoustic and optical branches. In the symmetric limit, $\gamma_e=\gamma_h\equiv\gamma$ and
$v_e=v_h\equiv v_{\rm th}$, the optical branch decouples exactly by symmetry, and one finds
\begin{equation}
\widetilde D_X(k,\omega)=D_X-
i\omega\,
\frac{2\gamma^2\bar v_X^2}{\gamma_X^2\left(\omega^2-v_{\rm th}^2k^2-i\gamma\omega\right)}.
\label{eq:Deff_sym}
\end{equation}
The onset of the instability can be estimated from Eq.~\eqref{eq:Deff_sym} 
by requiring $\mathrm{Im}(\omega)>0$, leading to the approximate condition $\gamma^2 \bar v_X^2 \gtrsim \gamma_X^2 v_{\rm th}^2$
in the long-wavelength limit. Therefore, the effective exciton diffusion coefficient is controlled by the
acoustic plasma response, while the optical plasmon remains present in the full secular equation but does not couple to the exciton mode in the perfectly symmetric case. Physically, this instability can be understood as a delayed feedback mechanism: 
density gradients in the exciton subsystem excite plasma waves, which in turn  induce currents that reinforce the original perturbation, leading to amplification instead of dissipation.

\section{Effective temperature of electron-hole plasma and excitons}
We now estimate the effective temperature of two sub-systems based on typical TMD experimental parameters. In analyzing the electron–hole plasma, we perform the estimates considering only electrons and assume that the hole temperature is the same. Such an approximation implies that electron and hole effective masses are taken to be equal \cite{Kormnyos2015}, and that their scattering times are also the same. We therefore use the relation $D_{X(e-h)}=k_{B}T_{X(e-h)}(\tau_{s}^{X(e-h)}/m_{X(e-h)})$ for the diffusion coefficient \cite{Mouri2014,Kulig2018}. Thus, the temperature (in eV) reads $k_{B}T_{X(e-h)}=D_{X(e-h)}m_{X(e-h)}/\tau_{s}^{X(e-h)}$. Typically, the exciton diffusion coefficient is about $D_{X}\sim 1-3$ $\rm cm^{2}/sec$, while experimental measurements for the electron–hole plasma give $D_{e-h}\sim10$ $\rm cm^{2}/sec$ \cite{Wagner2023,Zipfel2020,Wietek2024}. Exciton mass is between 0.5 and 1.2 of the free electron mass, and the electron effective mass is between 0.3 and 0.7 of the free electron mass depending on the material. We focus on $\rm MoSe_2$, for which $m_{X}=1.13m_{0}$ and $m_{e}=0.5m_{0}$. 

Finally, we extract from experimental measurements the scattering times for excitons and electrons. Typical scattering times for excitons vary from 30 fs to 1000 fs depending on the material and sample structure (monolayer or bi-layer) \cite{Zipfel2020,Wietek2024}. For the present estimates, we take $\tau_{X}\sim 100$ fs. For electrons in $\rm MoSe_2$, the scattering time was found to be about $\tau_{e}\sim70$ ps \cite{Venanzi2021}. Using these values, we estimate the temperatures of both sub-systems. For the excitonic sub-system $k_{B}T_{X}\sim6.4$ meV, while for the electron–hole plasma sub-system $k_{B}T_{e-h}\sim40$ $\mu$eV. The difference between the two sub-system temperatures therefore spans several orders of magnitude.

\section{Numerical simulations}

The analysis of exciton and electron–hole plasma expansion can be carried out both in free space (i.e., without geometrical constraints) and within a channel. In the former case, only quantitative differences in the exciton propagation speed between regimes are observed, while the shape of the exciton cloud remains qualitatively unchanged. In contrast, the channel geometry reveals a qualitative difference between propagation regimes. To validate the analytical predictions of the coupled exciton–plasma model, Eqs.~\eqref{eq_exciton} and \eqref{eq_plasma}, we perform time-dependent numerical simulations of the nonlinear diffusion dynamics in two spatial dimensions, considering a channel of length $L$ and width $w$, with $w \gg L$. The goal is to directly visualize the transition between the stable diffusive regime and the instability associated with an effective negative diffusion coefficient, as predicted by Eq.~\eqref{eq:Deff_full}. 

\begin{figure*}[t!]
\centering
\includegraphics[width=\linewidth]{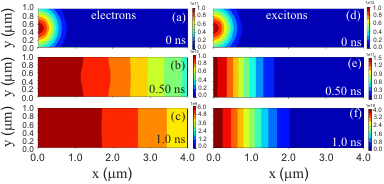}
\caption{Diffusion dynamics in a quasi-one-dimensional channel. 
Normal diffusive transport, corresponding to $\widetilde D_X>0$, where the exciton density profile broadens in time and the peak amplitude decreases, in agreement with the purely dissipative regime described by Eq.~\eqref{eq_diffusion_total}.}
\label{fig:positive diffusion}
\end{figure*}

\begin{figure*}[t!]
\centering
\includegraphics[width=\linewidth]{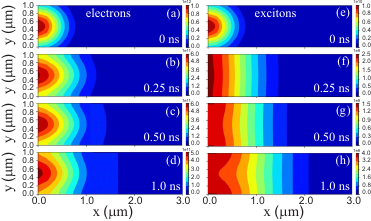}
\caption{Diffusion dynamics in a quasi-one-dimensional channel. 
Negative diffusion regime, corresponding to $\widetilde D_X<0$, where the exciton density develops a localized enhancement and increases in amplitude, signaling a dynamical instability. This behavior originates from the hybridization between the exciton diffusive mode and the acoustic plasma mode, as described by Eq.~\eqref{eq:Deff_full}, and provides a real-space manifestation of the gain predicted by the dispersion relation in Fig.~\ref{fig_dispersion}.}
\label{fig:negative diffusion}
\end{figure*}

The parameters used in the simulations are chosen to be consistent with experimentally reported values for monolayer TMDs (see Table~\ref{tab:parameters}). These values also agree with the estimates obtained in Sec.~V and satisfy the hierarchy required for the collisionless regime, $\gamma_X \gg \gamma$, ensuring that collective plasma effects dominate over diffusive relaxation.
\begin{table}[t!]
\label{tab:parameters}
\begin{ruledtabular}
\begin{tabular}{lcc}
Parameter & Value & Reference \\ \hline
Exciton diffusion coefficient $D_X$ 
& $\sim 3\,\mathrm{cm^2/s}$ 
& \cite{Moody2016Exciton,Ajayi2017} \\

Plasma thermal velocity $v_{\rm th}$ 
& $\sim 10^{5}\,\mathrm{m/s}$ 
& \cite{Raja2019} \\

Plasma frequency $\omega_p$ 
& $\sim 1$--$10\,\mathrm{THz}$ 
& \cite{Haug2009} \\

Exciton scattering time $\gamma_X^{-1}$ 
& $\sim 100\,\mathrm{fs}$ 
& \cite{Selig2016} \\

Carrier scattering time $\gamma^{-1}$ 
& $\sim 1$--$100\,\mathrm{ps}$ 
& \cite{Winzer2010,Brem2018} \\
\end{tabular}
\end{ruledtabular}
\caption{Parameters used in the numerical simulations, consistent with experimentally reported values for TMD monolayers.}
\end{table}
Figures~\ref{fig:positive diffusion}-\ref{fig:negative diffusion} illustrate two representative dynamical regimes obtained from Eqs.~\eqref{eq_exciton} and \eqref{eq_plasma}. Here, we present only the expansion dynamics of electrons and excitons, omitting the results for holes. This is justified by the symmetry between electrons and holes: they share identical effective masses and equal initial concentrations, leading to nearly identical time evolution of their densities.

\paragraph{(i) Positive diffusion regime.}

For weak exciton--plasma coupling (small $\gamma_\alpha/\gamma_X$), the effective diffusion coefficient remains positive, $\widetilde D_X>0$. In this regime, the exciton density profile exhibits the standard Gaussian broadening expected from Eq.~\eqref{eq_diffusion_total}, with the peak density decreasing monotonically in time (see Fig.~\ref{fig:positive diffusion} for details). This behavior is consistent with the purely dissipative dispersion relation obtained in Eq.~\eqref{eq_disp1}.

\paragraph{(ii) Negative diffusion regime.}

For sufficiently strong coupling or in the presence of resonant conditions with the acoustic plasma mode, Eq.~\eqref{eq:Deff_sym} predicts that $\widetilde D_X$ acquires an imaginary contribution with opposite sign, leading to $\mathrm{Im}(\omega)>0$. In real space, this results in a strikingly different behavior: instead of broadening, the exciton density develops a localized peak that grows in amplitude over time, signaling a dynamical instability. This effect is clearly visible in Fig.~\ref{fig:negative diffusion}, where the exciton cloud undergoes self-focusing and density accumulation at the center. Such behavior corresponds to the regime in which the exciton diffusive mode hybridizes with the acoustic plasma branch, as discussed in Sec.~IV.

The numerical results confirm that the sign of the effective diffusivity is controlled by the competition between two mechanisms already identified analytically: i) Dissipative relaxation governed by $D_X$ in Eq.~\eqref{eq_diffusion_total}; 
ii) Reactive (inertial) feedback induced by plasma oscillations, encoded in Eq.~\eqref{eq:Deff_full}. When the second mechanism dominates, density fluctuations are amplified rather than damped, leading to a regime of effective negative diffusion of excitons. Importantly, this transition does not require any nonlinear modification of the diffusion law, but emerges naturally from the linear coupling to collective plasma modes.

The simulated dynamics are qualitatively consistent with recent experimental observations of anomalous exciton transport in TMD monolayers, where spatial contraction or non-monotonic expansion of exciton clouds has been reported \cite{Wietek2024,Zipfel2020,Kumar2014}. In particular, the emergence of localized density enhancements can be interpreted as a manifestation of the instability predicted by Eq.~\eqref{eq:Deff_full}, providing further support for the role of electron--hole plasma dynamics in exciton transport.

\section{Conclusion}

We have developed a minimal hydrodynamic framework to describe exciton transport in the presence of an electron--hole plasma in two-dimensional semiconductors. Treating excitons, electrons, and holes as coupled fluids, we have shown that momentum exchange with the plasma provides a natural and unified mechanism for the strong renormalization of exciton diffusion observed in experiments.

In the collisional, quasi-neutral regime, mutual diffusion between excitons and plasma leads to a redistribution of transport coefficients while preserving the positivity of the effective exciton diffusivity. By contrast, when plasma inertia and collective charge dynamics are taken into account, the exciton diffusive mode hybridizes with acoustic plasma excitations. This coupling gives rise to a dynamical instability, manifested as an effective negative diffusion coefficient. Importantly, the instability does not rely on nonlinear diffusion laws or thermodynamic anomalies, but instead emerges from the non-equilibrium coupling between slow excitonic motion and fast plasma degrees of freedom.

A key experimental fingerprint of the mechanism proposed here is the transient spatial contraction of the exciton cloud at early times, followed by re-expansion once the plasma sufficiently relaxes. Such non-monotonic dynamics is qualitatively distinct from the scenario proposed by Wietek \emph{et al.}~\cite{Wietek2024}, where negative diffusivity originates from a nonlinear coupling between a rapidly expanding electron--hole plasma and slower excitonic populations across the Mott threshold. In that picture, the effective diffusion coefficient becomes density dependent and the sign reversal is transient and spatially homogeneous. In contrast, the present mechanism predicts a fundamentally \emph{wavevector-selective} instability: density fluctuations satisfying
\[
\gamma^2 v_X^2 \gtrsim \gamma_X^2\left(v_{\rm th}^2k^2+\omega_p^2/k^2\right)
\]
are amplified, whereas longer- and shorter-wavelength modes remain stable. Such momentum selectivity should be experimentally accessible using spatially resolved pump--probe or transient-grating spectroscopy in highly excited TMD monolayers.

Our results therefore provide a transparent physical interpretation of recent observations of effective negative exciton diffusion in multicomponent systems, while clarifying the distinct roles played by collisional transport and collective plasma dynamics. More broadly, they highlight the importance of treating excitons as part of an interacting many-fluid system rather than as isolated diffusive quasiparticles.

The present work also calls attention to the role of \emph{collective} plasma excitations, rather than individual carrier scattering events, as a dominant control parameter of exciton transport. While exciton--phonon and exciton--exciton  interactions have been extensively investigated, the coupling between excitonic density fluctuations and the acoustic branch of the electron--hole plasma has so far remained largely unexplored. Since the present framework can be naturally extended to include disorder, finite recombination, and external driving fields, it may serve as a starting point for a unified description of the broad phenomenology of exciton transport in two-dimensional materials, including recently observed superfluid-like propagation~\cite{delAguila2023} and viscous hydrodynamic flow~\cite{Mantsevich2024}.

\section{Acknowledgements}

V.N.M. thanks for support Russian Science Foundation grant No. 24-12-00020.

\end{document}